\documentclass[pdflatex,sn-mathphys-num]{sn-jnl}% Math and Physical Sciences Numbered Reference Style 
\usepackage{graphicx}%
\usepackage{multirow}%
\usepackage{amsmath,amssymb,amsfonts}%
\usepackage{amsthm}%
\usepackage{mathrsfs}%
\usepackage[title]{appendix}%
\usepackage{xcolor}%
\usepackage{textcomp}%
\usepackage{manyfoot}%
\usepackage{booktabs}%
\usepackage{algorithm}%
\usepackage{algorithmicx}%
\usepackage{algpseudocode}%
\usepackage{listings}%
\usepackage{graphicx}% Include figure files
\usepackage[switch,pagewise,running]{lineno}
\usepackage{dcolumn}% Align table columns on decimal point
\usepackage{bm}% bold math
\usepackage{epsfig}
\usepackage{xspace}
\usepackage{hyperref}
\hypersetup{colorlinks,linkcolor=red,citecolor=blue,urlcolor=blue}  
\usepackage{color,xcolor}
\usepackage{verbatim}
\usepackage{float}
\usepackage{slashed}
\usepackage{rotating}
\usepackage[normalem]{ulem} 
\usepackage[caption=false]{subfig}
\usepackage{amsmath}
\theoremstyle{thmstyleone}%
\theoremstyle{thmstyletwo}%

\theoremstyle{thmstylethree}%

\begin{document}

\title[Article Title]{Time-Dependent Leptohadronic Modeling of Markarian 421}


\author*[1]{\fnm{Rui} \sur{Xue}}\email{ruixue@zjnu.edu.cn}

\author*[2]{\fnm{Ze-Rui} \sur{Wang}}\email{zerui\_wang62@163.com}
%equalcont{These authors contributed equally to this work.}

\author[1]{\fnm{Hai-Bin} \sur{Hu}}%\email{iiiauthor@gmail.com}

\affil*[1]{\orgdiv{Department of Physics}, \orgname{Zhejiang Normal University}, \orgaddress{\city{Jinhua}, \postcode{321004}, \country{China}}}

\affil*[2]{\orgdiv{College of Physics and Electronic Engineering}, \orgname{Qilu Normal University}, \orgaddress{\city{Jinan}, \postcode{250200}, \country{China}}}


\abstract{Due to its proximity, Markarian 421 is one of the most extensively studied jetted active galactic nuclei. Its spectral energy distribution and light curve are widely studied, serving as primary means for understanding jet radiation mechanisms. Numerous intriguing observational results have been discovered, some of which, such as the hard X-ray excess, and the associated variability between X-ray and very-high-energy (VHE) emissions, challenge the commonly adopted one-zone leptonic model. In this work, by establishing a time-dependent leptohadronic model, we explore whether the hard X-ray excess and the associated variability between X-ray and VHE emissions could be interpreted by emission from hadronic interactions. Our modeling finds that for the hard X-ray excess found in 2013, both of the secondary emissions from photohadronic and hadronuclear interactions could be a possible explanation for the hard X-ray excess without introducing a super-Eddington jet power. The emission from the photohadronic interactions contributes only to the hard X-ray band, while the hadronuclear interactions also predict VHE emissions associated with the hard X-rays. While for the hard X-ray excess found in 2016, only the secondary emissions from photohadronic interactions provide an interpretation at the cost of introducing a super-Eddington jet power. For the associated variability between X-ray and VHE emissions in 2017, we find that hadronic interactions fail to provide a possible interpretation.}

\maketitle

\section{Introduction}\label{sec1}

Blazar is the highly variable subclass of active galactic nuclei (AGNs) with relativistic jets pointing along the observers' line of sight \citep{1995PASP..107..803U}. Its electromagnetic emission exhibits wide energy range spectral energy distributions (SEDs) characterized by two bumps \citep{1997ARA&A..35..445U, 2019NewAR..8701541H, 2024NewAR..9801693C}. The low-energy bump is attributed to synchrotron radiation from primary relativistic electrons in the jet \citep{1997A&A...325..109S}. However, the origin of the high-energy bump remains a topic of debate. In leptonic models, this bump results from inverse Compton (IC) radiation, where relativistic electrons scatter soft photons emitted by the same electron population (synchrotron-self Compton, SSC; \cite{1985ApJ...298..114M,1992ApJ...397L...5M,1992A&A...256L..27D}), or soft photons from fields outside jet (external Compton) like the accretion disk \citep{1993ApJ...416..458D}, the broad-line region (BLR; \citep{1994ApJ...421..153S}), or the dusty torus \citep{2000ApJ...545..107B}. In recent years, an increasing number of blazars have been found to potentially correlate with high-energy neutrinos (e.g., \citep{2018Sci...361.1378I, 2018Sci...361..147I, 2022Sci...378..538I, 2017PhRvD..96h2001A, 2020PhRvL.124e1103A, 2020ApJ...892...92A}). As a result, hadronic models have come into focus and are widely discussed. In hadronic models, the high-energy bump may arise from proton synchrotron radiation \citep{2000NewA....5..377A, 2003APh....18..593M, 2023PhRvD.107j3019X}, and the emission from secondary particles that are generated in the hadronic interactions and internal $\gamma \gamma$ pair production \citep{2013MNRAS.434.2684M, 2017MNRAS.464.2213P, 2022A&A...659A.184L, 2022PhRvD.106j3021X, 2024A&A...685A.110P, 2024ApJS..271...10W}.

Markarian 421 (Mrk 421; RA=11$^{\rm h}$04$^{\rm m}$27$^{\rm s}$, Dec=+38$^\circ$$12^\prime$$32^{\prime \prime}$, J2000), located at a redshift $z = 0.031$ \citep{1975ApJ...198..261U}, is the first extragalactic very-high-energy (VHE; 0.1--100 TeV) emitter detected by the Whipple telescope in 1992 \citep{1992Natur.358..477P}. According to its peak frequency of the low-energy bump $\nu^{\rm s}_{\rm p}$ and feature of weak emission lines, Mrk 421 is classified into the high synchrotron peaked BL Lac (HBL, $\nu^{\rm s}_{\rm p}\gtrsim10^{15}~\rm Hz$; \citep{1995PASP..107..803U, 2010ApJ...716...30A}), which is generally the most nearby representative and brightest extragalactic $\gamma$-ray source in the TeV sky. Due to the scarcity of external photon fields, which is also a characteristic of HBLs \citep{1995PASP..107..803U}, the SED of Mrk 421 is usually explained using the SSC model within the framework of the classic one-zone model \citep{2011ApJ...736..131A}. However, it faces challenges in explaining some unique features of the SED. For example, a significant hard X-ray excess above $20\rm~keV$ during low state of Mrk 421 had been reported in 2013 \cite{2016ApJ...827...55K} and 2016 \citep{2021MNRAS.504.1427A}, respectively. Due to constraint of the radio data, Chen \cite{2017ApJ...842..129C} argued that the excess in 2013 cannot be interpreted by the low-energy tail of the SSC emission in the framework of one-zone leptonic model. To interpret this excess, Chen \cite{2017ApJ...842..129C} applied the spine-layer (i.e., two-zone) model, Hu et al. \cite{2023ApJ...948...82H} introduced a second injection of relativistic electrons that accelerated by stochastic acceleration, and Xue et al. \cite{2023PhRvD.107j3019X} attempted the one-zone proton synchrotron model. For the time-dependent modeling of one-zone leptonic model, it naturally predicts inherent correlations between multiwavelength light curves, as the full band emission is contributed by the same population of relativistic electrons. Such correlations have indeed been widely discovered in multiwavelength observations, supporting the notion of a simple one-emitting-zone origin to some extent (e.g., \citep{2016A&A...593A..91A, 2016ApJ...819..156B}). However, more and more multiwavelength observations indicate that the variability patterns of Mrk 421 are complex and challenging to explain using a one-zone SSC model. The correlation between X-ray and VHE variabilities has been found in flare (e.g., \citep{2008ApJ...677..906F, 2011ApJ...738...25A, 2013PASJ...65..109C, 2015A&A...578A..22A}) and low activity \citep{2015A&A...576A.126A, 2016ApJ...819..156B} states widely, however, the other bands' activities are found to have weak or no correlation to those of X-ray and VHE bands \citep{1995ApJ...449L..99M, 2013PASJ...65..109C, 2007ApJ...663..125A, 2016ApJ...819..156B, 2021MNRAS.504.1427A}. Recently reported detailed multiwavelength light curves of a flare are difficult to explain with the one-zone SSC model \citep{2020ApJ...890...97A}, therefore two-zone \citep{2021MNRAS.505.2712D} and multi-zone \citep{2019MNRAS.487..845B} time-dependent leptonic models are introduced. Some studies have found that when X-ray and VHE emissions exhibit coordinated flares, the optical/ultraviolet and MeV--GeV emissions are generally in a low or inactive state \citep{2015A&A...578A..22A, 2021A&A...655A..89M}. Such multiwavelength characteristics are challenging to explain using a one-zone SSC model. Currently, a commonly adopted approach is the two-zone SSC model, where one zone produces low-state SED (similar to the traditional one-zone model), while another region with a narrow electron distribution which may arise through stochastic acceleration \cite{2014ApJ...780...64A} and magnetic reconnection \citep{2020ApJS..248...29A} generates the X-ray and VHE flares.

Due to the discovery of potential association events between AGNs and high-energy neutrinos, interest in hadronic models has been rekindled. In some case studies, it has been found that secondary emission from hadronic processes can contribute potentially, or even dominate, the X-ray emission \cite{2013MNRAS.434.2684M, 2015MNRAS.447...36P, 2019MNRAS.483L..12C, 2019NatAs...3...88G, 2022PhRvD.106j3021X}, therefore the hard X-ray excess might be indicative of the emergence of hadronic radiation features. Meanwhile, the energetic electrons that generate hard X-ray excess may also contribute at VHE band through IC scattering. Other bands emission remains primarily governed by the leptonic emission from primary electrons. As a result the flares in X-ray and VHE bands naturally lack correlations with other bands dominated by leptonic processes. Therefore, the hadronic model could be a plausible explanation for the hard X-ray excess and variability patterns observed in Mrk 421. In this work, we establish a time-dependent leptohadronic model to interpret the hard X-ray excess, and test if the correlated X-ray and VHE flares observed in 2017 February 4 \citep{2021A&A...655A..89M} could originate from hadronic interactions. This paper is structured as follows. In Sec.~\ref{model}, the description of the time-dependent leptohadronic model is presented. In Sec.~\ref{app}, we apply the model to explain two observations of hard X-ray excesses. In Sec.~\ref{correlation}, we test if hadronic interactions can interpret the flaring state SED of Mrk 421 in 2017. We end with discussion and conclusion in Sec.~\ref{DC}. Throughout the paper, the cosmological parameters $H_{0}=69.6\ \rm km\ s^{-1}Mpc^{-1}$, $\Omega_{0}=0.29$, and $\Omega_{\Lambda}$= 0.71 \cite{2014ApJ...794..135B} are adopted.

\section{Model description}\label{model}
Following the conventional one-zone jet model (e.g., \citep{2001A&A...367..809K}), we assume that all the blazar's jet emission is predominantly contributed by one emitting region. This emitting region is assumed to be a spherical blob composed of a plasma of charged particles in a uniformly entangled magnetic field $B$ with radius $R$ and moving with the bulk Lorentz factor $\Gamma=(1-\beta^2)^{-1/2}$, where $\beta c$ is the speed of the blob, at a viewing angle with respect to observers' line of sight. The blob is assumed to be a homogeneous emitting region, with charged particles and photons uniformly and isotropically distributed. For the relativistic jet close to the line of sight in blazars with a viewing angle of $\theta \lesssim 1/\Gamma$, we have the Doppler factor $\delta_{\rm D} \approx \Gamma$. Due to the beaming effect, all emissions from the emitting blob will be amplified by a factor $\delta_{\rm D}$ in energy and a factor $\delta_{\rm D}^4$ in luminosity, and the observed timescale will be shortened by a factor $\delta_{\rm D}^{-1}$. In the following, all parameters are measured in the comoving frame, unless specified otherwise.

Through various acceleration mechanisms \citep{2020LRCA....6....1M}, primary relativistic electrons with a broken power-law distribution 
%\footnote{In general, conventional acceleration mechanisms predict a power-law distribution for the accelerated particles. Then, forming a broken power-law distribution under the influence of radiative cooling. However, the observed index difference is frequently significantly larger than 0.5 \citep{2012ApJ...753..154M, 2022MNRAS.514.3074B}, which does not align with the inference of radiative cooling \cite{1979rpa..book.....R}. In this work, we simply assume a broken power-law distribution for the injected relativistic electrons as suggested in \citep{2009MNRAS.397..985G, 2010MNRAS.402..497G}, so that the SED fitting would be much easier. For a possible physical origin of the broken power-law electron distribution, one may refer to \cite{2022MNRAS.514.3074B, 2024MNRAS.529..903T}.}
, 
\begin{equation}
\dot{Q}^{\rm inj}_{\rm e}(\gamma_{\rm e}, t)=\dot{Q}_{\rm e,0}\gamma_{\rm e}^{-\alpha_{\rm e,1}}[1+(\frac{\gamma_{\rm e}}{\gamma_{\rm e,b}})^{(\alpha_{\rm e,2}-\alpha_{\rm e,1})}]^{-1} \times \Theta(t_{\rm inj}-t), \gamma_{\rm e,min}<\gamma_{\rm e}<\gamma_{\rm e,max},
\end{equation}
and primary relativistic protons with a power-law distribution
\begin{equation}
    \dot{Q}^{\rm inj}_{\rm p}(\gamma_{\rm p}, t)=\dot{Q}_{\rm p,0}\gamma_{\rm p}^{-\alpha_{\rm p}}\Theta(t_{\rm inj}-t), \gamma_{\rm p,min}<\gamma_{\rm p}<\gamma_{\rm p,max},
\end{equation}
at constant rates are continuously injected into the blob for the injection time $t_{\rm inj}$. Here, subscripts `e/p' represent quantities for electrons or protons, $\dot{Q}_{\rm e/p,0}$ are normalizations in units of $\rm cm^{-3}~s^{-1}$, $\gamma_{\rm e,min/b/max}$ are the minimum, break, and maximum electron Lorentz factors, $\alpha_{\rm e,1}$, $\alpha_{\rm e,2}$ are the electron spectral indices before and after $\gamma_{\rm e,b}$, $\gamma_{\rm p,min/max}$ are the minimum, and maximum proton Lorentz factors, $\alpha_{\rm p}$ is the proton spectral index. By giving an electron/proton injection luminosity $L_{\rm e/p,inj}$, $\dot{Q}_{\rm e/p,0}$ can be determined by $\int \gamma_{\rm e/p}m_{\rm e/p}c^2 \dot{Q}^{\rm inj}_{\rm e/p}(\gamma_{\rm e/p}, t) d\gamma_{\rm e/p}=\frac{3}{4\pi R^3}L_{\rm e/p,inj}$, where $m_{\rm e/p}$ is the particle rest mass. For $\Theta(t_{\rm inj}-t)$, it represents the Heaviside step function and means that the continuous injection of relativistic particles stops at $t_{\rm inj}$. The temporal evolution of the primary charged particles energy distribution $N_{\rm e/p}(\gamma_{\rm e/p}, t)~\rm [cm^{-3}]$ is governed by the continuity equation
\begin{equation}\label{kprimary}
\frac{\partial N_{\rm e/p}(\gamma_{\rm e/p}, t)}{\partial t}=\frac{\partial}{\partial \gamma_{\rm e/p}}[\dot{\gamma}_{\rm e/p}N_{\rm e/p}(\gamma_{\rm e/p}, t)]+\dot{Q}^{\rm inj}_{\rm e/p}(\gamma_{\rm e/p}, t)-\frac{N_{\rm e/p}(\gamma_{\rm e/p}, t)}{t_{\rm esc}},
\end{equation}
where $\dot{\gamma}_{\rm e/p}$ is the cooling rate for primary charged particles, and $t_{\rm esc}=R/c$ is the escape timescale. More specifically, $\dot{\gamma}_{\rm e}$ considering synchrotron and IC cooling for electrons is given by
\begin{equation}
\dot{\gamma}_{\rm e} = \frac{4\sigma_{\rm T}}{3m_{\rm e}c}[U_{\rm B}+\kappa_{\rm KN}U_{\rm soft}]\gamma_{\rm e}^2,
\end{equation}
where $\sigma_{\rm T}$ is the Thomson cross-section, $U_{\rm B}=B^2/8\pi$ is the energy density of the magnetic field, $U_{\rm soft}$ is the energy density of the soft photons from the primary and secondary particles, and 
\begin{equation}
\kappa_{\rm KN}=\frac{9}{U_{\rm soft}}\int^{\infty}_0dE_{\rm soft}~E_{\rm soft}~n_{\rm soft}(E_{\rm soft},t)\times \int^1_0 dq\frac{2q^2\textmd{ln}~q+q(1+2q)(1-q)+\frac{q(\omega q)^2(1-q)}{2(1+\omega q)}}{(1+\omega q)^3}
\end{equation}
is a numerical factor accounting for the Klein-Nishina effect \citep{2010NJPh...12c3044S}, where $E_{\rm soft}$ is the energy of soft photons, $n_{\rm soft}(E_{\rm soft},t)$ is the number density distribution of soft photons emitted by the primary electrons and secondary pairs (how it is derived will be given in Eq.~\ref{kphoton}), $\omega=4E_{\rm soft} \gamma_{\rm e}/(m_{\rm e}c^2)$. For relativistic protons, its cooling rate $\dot{\gamma}_{\rm p}$ can be written as
\begin{equation}
\dot{\gamma}_{\rm p} = \dot{\gamma}_{\rm p}^{\rm syn} + \dot{\gamma}_{\rm p}^{p\gamma}+\dot{\gamma}_{\rm p}^{\rm BH}+\dot{\gamma}_{\rm p}^{pp},
\end{equation}
where the four terms on the right-hand side describe rates of synchrotron, photopion ($p\gamma$) interaction, Bethe--Heitler (BH) pair production, and hadronuclear ($pp$) interaction, respectively. These cooling rates are given by
\begin{equation}
\begin{split}
&\dot{\gamma}_{\rm p}^{\rm syn} = \frac{4\sigma_{\rm T}U_{\rm B}}{3m_{\rm e}c}(\frac{m_{\rm e}}{m_{\rm p}})^3\gamma_{\rm p}^2\\
&\dot{\gamma}_{\rm p}^{p\gamma} = \frac{c}{2\gamma_{\rm p}} \int_{E_{\rm th}^{p\gamma}/{2\gamma_{\rm p}}}^{\infty} dE_{\rm soft} \frac{n_{\rm soft}(E_{\rm soft},t)}{E_{\rm soft}^2} \int_{E_{\rm th}^{p\gamma}}^{2E\gamma_{p}} dE_{r} \times \sigma_{p\gamma}(E_{r})K_{p\gamma}(E_{r})E_{r}\\
&\dot{\gamma}_{\rm p}^{\rm BH} = \frac{c}{2\gamma_{\rm p}} \int_{E_{\rm th}^{\rm BH}/{2\gamma_{\rm p}}}^{\infty} dE_{\rm soft} \frac{n_{\rm soft}(E_{\rm soft},t)}{E_{\rm soft}^2} \int_{E_{\rm th}^{\rm BH}}^{2E\gamma_{p}} dE_{r}\times \sigma_{\rm BH}(E_{r})K_{\rm BH}(E_{r})E_{r}\\
&\dot{\gamma}_{\rm p}^{\rm pp} = \gamma_{\rm p}cn_{\rm H}K_{\rm pp}\sigma_{\rm pp}(\gamma_{\rm p}),
\end{split}
\end{equation}
respectively, where $E_{\rm th}^{p\gamma}=145~\rm MeV$ and $E_{\rm th}^{\rm BH}=1~\rm MeV$ are the photon threshold energies in the rest frame of protons for the $p\gamma$ and BH processes \citep{2008PhRvD..78c4013K}, $E_{r}$ is the photon energy in the rest frame of protons, $\sigma_{\rm p\gamma/BH}$ are cross sections for $p\gamma$ \citep{1990ApJ...362...38B, 2000CoPhC.124..290M} and BH processes \citep{1992ApJ...400..181C}, $K_{\rm p\gamma/BH}$ are the inelasticities of $p\gamma$ \citep{1990ApJ...362...38B, 2000CoPhC.124..290M} and BH processes \citep{1992ApJ...400..181C}, $n_{\rm H}$ is the number density of cold protons, $K_{\rm pp}\approx 0.5$ is the inelasticity coefficient, and $\sigma_{\rm pp}$ is the cross section for inelastic $pp$ interactions \citep{2006PhRvD..74c4018K}.

In addition to emissions from primary electrons and protons, secondary electrons/positrons (pairs) generated in hadronic interactions and internal $\gamma \gamma$ annihilation also would generate significant emissions. The temporal evolution of the secondary pairs energy distribution $N_{\rm e,sec}(\gamma_{\rm e}, t)~\rm [cm^{-3}]$ is governed by the continuity equation similar to Eq.~(\ref{kprimary}), i.e.,
\begin{equation}\label{ksecondary}
\frac{\partial N_{\rm e,sec}(\gamma_{\rm e}, t)}{\partial t}=\frac{\partial}{\partial \gamma_{\rm e}}[\dot{\gamma}_{\rm e}N_{\rm e,sec}(\gamma_{\rm e}, t)]-\frac{N_{\rm e,sec}(\gamma_{\rm e}, t)}{t_{\rm esc}}+\dot{Q}^{p\gamma}_{\rm e}(\gamma_{\rm e}, t)+\dot{Q}^{\rm BH}_{\rm e}(\gamma_{\rm e}, t)+\dot{Q}^{pp}_{\rm e}(\gamma_{\rm e}, t)+\dot{Q}^{\gamma \gamma}_{\rm e}(\gamma_{\rm e}, t).
\end{equation} 
The only difference is replacing the injection of primary electrons $\dot{Q}^{\rm inj}_{\rm e}(\gamma_{\rm e}, t)$ with $\dot{Q}^{p\gamma}_{\rm e}(\gamma_{\rm e}, t)+\dot{Q}^{\rm BH}_{\rm e}(\gamma_{\rm e}, t)+\dot{Q}^{pp}_{\rm e}(\gamma_{\rm e}, t)+\dot{Q}^{\gamma \gamma}_{\rm e}(\gamma_{\rm e}, t)$, which represents differential spectrum of secondary pairs from $p\gamma$, BH, $pp$, and $\gamma \gamma$ productions. In the calculation, $\dot{Q}^{p\gamma}_{\rm e}(\gamma_{\rm e}, t)$, $\dot{Q}^{\rm BH}_{\rm e}(\gamma_{\rm e}, t)$, and $\dot{Q}^{pp}_{\rm e}(\gamma_{\rm e}, t)$ are obtained with analytical expressions developed in Refs. \citep{2006PhRvD..74c4018K, 2008PhRvD..78c4013K}, and the expression of $\dot{Q}^{\gamma \gamma}_{\rm e}(\gamma_{\rm e}, t)$ is given by \citep{1983Afz....19..323A}
\begin{equation}
   \begin{split}
  & \dot{Q}^{\gamma \gamma}_{\rm e}(\gamma_{\rm e}, t) =  \frac{3\sigma_{\rm T} c}{16}\ \int_{\gamma_{\rm e}}^{\infty}{dE_{\gamma}\ \frac{n_{\gamma}(E_\gamma,t)}{\epsilon_{\gamma}^3}}\int_{\frac{\epsilon_{\gamma}}{4\gamma_{\rm e}(\epsilon_{\gamma}-\gamma_{\rm e})}}^{\infty}{dE_{\rm soft}}\ \frac{n_{\rm soft}(E_{\rm soft},t)}{\epsilon_{\rm soft}^2} \left[ \frac{4\epsilon_{\gamma}^2}{\gamma_{\rm e}(\epsilon_{\gamma}-\gamma_{\rm e})} \right. \\
  & \left. \times \ln{\left( \frac{4\gamma_{\rm e} \epsilon_{\rm soft} (\epsilon_{\gamma}-\gamma_{\rm e})}{\epsilon_{\gamma}} \right)} -8\epsilon_{\gamma} \epsilon_{\rm soft}+\frac{2\epsilon_{\gamma}^2(2\epsilon_{\gamma}\epsilon_{\rm soft}-1)}{\gamma_{\rm e}(\epsilon_{\gamma}-\gamma_{\rm e})} -\left(1-\frac{1}{\epsilon_{\gamma}\epsilon_{\rm soft}}\right)\left( \frac{\epsilon_{\gamma}^2}{\gamma_{\rm e}(\epsilon_{\gamma}-\gamma_{\rm e})} \right)^2 \right],
   \end{split}
   \end{equation}
where $\epsilon_{\rm soft}$ and $\epsilon_{\gamma}$ are the dimensionless energies of low- and high-energy photons, and $n_{\gamma}(E_\gamma,t)$ is the number density distribution of high-energy photons. In the time-dependent modeling, the expression of $n_{\rm soft/\gamma}(E_{\rm soft/\gamma},t)$ is obtained by solving the continuity equation
\begin{equation}\label{kphoton}
\frac{\partial n_{\rm soft/\gamma}(E_{\rm soft/\gamma},t)}{\partial t}=\dot{Q}_{\rm soft/\gamma}(E_{\rm soft/\gamma},t)-\frac{n_{\rm soft/\gamma}(E_{\rm soft/\gamma},t)}{t_{\rm soft/\gamma}^{\rm abs}}-\frac{n_{\rm soft/\gamma}(E_{\rm soft/\gamma},t)}{t_{\rm esc}},
\end{equation}
where $\dot{Q}_{\rm soft/\gamma}(E_{\rm soft/\gamma},t)$ represents the differential spectra of synchrotron and IC emissions from primary electrons and secondary pairs and emissions from $\pi^0$ decay. In our calculation, the public NAIMA python package\footnote{https://naima.readthedocs.io/en/latest/} is applied to calculate the leptonic emission from primary and secondary pairs \citep{2015ICRC...34..922Z}, and the differential spectrum of $\pi^0$ decay photons in $pp$ interactions. For the differential spectrum of $\pi^0$ decay photons in $p\gamma$ interactions, we apply the analytical expressions developed in \cite{2008PhRvD..78c4013K}. The generated soft and high-energy photons would be absorbed through distinct effects. More specifically, for soft photons, synchrotron self-absorption is important. Here, 
\begin{equation}
t_{\rm soft}^{\rm abs} = -\big[\frac{c}{8\pi \nu^2 m_{\rm e}} \int d\gamma_{\rm e} P_{\rm e}(\nu, \gamma_{\rm e})\gamma^2_{\rm e} \frac{\partial}{\partial \gamma_{\rm e}}[\frac{N_{\rm e}(\gamma_{\rm e},t)}{\gamma_{\rm e}^2}] \big]^{-1}
\end{equation}
represents the timescale of synchrotron self-absorption \citep{1979rpa..book.....R}, where $\nu$ is the photon frequency, and $P_{\rm e}(\nu, \gamma_{\rm e})$ is the synchrotron emission coefficient for a single electron integrated over the isotropic distribution of pitch angles \citep{2010PhRvD..82d3002A}, and for high-energy $\gamma$-ray photons $\gamma \gamma$ annihilation could be significant. Here,
\begin{equation}
t_{\gamma}^{\rm abs} =\big[\frac{2c(m_{\rm e}c^2)^4}{E^2_\gamma}\int_{m_{\rm e}^2c^4/E_\gamma}^{\infty} d\epsilon \frac{n_{\rm soft}(E_{\rm soft})}{E_{\rm soft}^2}\int_1^{\frac{E_{\rm soft}E_\gamma}{m_{\rm e}^2c^4}}ds\times \sigma_{\rm \gamma\gamma}(s)s\big]^{-1}
\end{equation}
is the timescale of $\gamma \gamma$ annihilation, where $\sigma_{\rm \gamma\gamma}$ is the $\gamma\gamma$ pair-production cross section, and $\sqrt{s}$ is the center-of-momentum frame Lorentz factor of the produced pairs \citep{2009herb.book.....D}. %Please note that in Eqs.~(\ref{kprimary}, \ref{ksecondary}, \ref{kphoton}), we assume that charged particles and photons have the same escape timescale, i.e., $R/c$, which is an oversimplified assumption. This implies that changes in the photon luminosity will closely track changes in the injection rate, with any time lags between light curves depending on different cooling times \citep{1999MNRAS.306..551C, 2008A&A...491L..37M, 2013MNRAS.434.2684M, 2024A&A...685A.110P}. When considering a detailed acceleration process, the escape timescale of particles is likely to be energy-dependent due to the energy-dependent diffusion (e.g., \citep{2011ApJ...739...66T}).

To study the temporal evolution of primary electrons and protons (Eq.~\ref{kprimary}), secondary pairs (Eq.~\ref{ksecondary}) and photons (Eq.~\ref{kphoton}), the fully implicit difference scheme is adopted here to solve continuity equations \citep{1970JCoPh...6....1C}. Here we take Eq.~(\ref{kprimary}) as an example. We define the energy mesh points of relativistic particles with logarithmic steps $\gamma_j = \gamma_{\rm min}\left( \frac{\gamma_{\rm max}}{\gamma_{\rm min}} \right)^\frac{j-1}{j_{\rm max}-1}$, where $j_{\rm max}$ is the mesh points number, and the energy intervals are $\Delta \gamma_j = \gamma_{j+1/2} -\gamma_{j-1/2}$. In our calculation, a grid of 100 points has been used both for particle energy and photon frequency. Quantities with the subscript $j\pm 1/2$ are calculated at half grid points. To discretize the continuity equation, we define $N^i_j = N(\gamma_j, i\Delta t)$ and $F^{i+1}_{j\pm 1/2} = \dot\gamma^i_{j \pm 1/2} N^{i+1}_{j\pm 1/2}$. Then Eq.~(\ref{kprimary}) becomes $\frac{N^{i+1}_j - N^i_j}{\Delta t} = \frac{F^{i+1}_{j+1/2} - F^{i+1}_{j-1/2}}{\Delta \gamma} + Q^i_j - \frac{N^{i+1}_j}{t_{\rm esc}}$. In this specific case, we have $N_{j+1/2}\equiv N_{j+1}$ and $N_{j-1/2}\equiv N_{j}$, according to the prescriptions of \cite{1970JCoPh...6....1C}. By discretizing the continuity equations in time $i$ and energy $j$, Eq.~(\ref{kprimary}) can be rewritten as \citep{1999MNRAS.306..551C}
\begin{equation}
V3_jN^{i+1}_{j+1}+V2_jN^{i+1}_j+V1_jN^{i+1}_{j-1}=S^i_j,
\end{equation}
where
\begin{equation}
\begin{split}
&V1_j=0,\\ 
&V2_j=1+\frac{\Delta t}{\gamma_{j+1/2}-\gamma_{j-1/2}}\dot{\gamma}_{j-1/2}^i+\frac{\Delta t}{t_{\rm esc}},\\
&V3_j=-\frac{\Delta t}{\gamma_{j+1/2}-\gamma_{j-1/2}}\dot{\gamma}_{j+1/2}^i,\\
&S^i_j=N^i_j+Q^i_j\Delta t. 
\end{split}
\end{equation}
Then the discretized equation forms a tridiagonal matrix
%\begin{widetext}
\begin{eqnarray}\label{matrix}
		\left(
		\begin{array}{cccccc}
			V2_0&V3_0&0&.&.&0\\
			V1_1&V2_1&V3_1&0&.&.\\
			.&.&.&.&.&.\\
			.&.&.&.&.&.\\
			.&.&.&.&.&.\\
			.&.&0&V1_{j_{\rm max}-1}&V2_{j_{\rm max}-1}&V3_{j_{\rm max}-1}\\
			.&.&0&0&V1_{j_{\rm max}}&V2_{j_{\rm max}}\\
		\end{array}
		\right)\left(
		\begin{array}{c}
		N_0^{i+1}\\
		N_1^{i+1}\\
		.\\
		.\\
		.\\
		N_{j_{\rm max}-1}^{i+1}\\
		N_{j_{\rm max}}^{i+1}\\
		\end{array}
		\right)=\left(
		\begin{array}{c}
		S_0^{i}\\
		S_1^{i}\\
		.\\
		.\\
		.\\
		S_{j_{\rm max}-1}^{i}\\
		S_{j_{\rm max}}^{i}\\
		\end{array}
		\right),
	\end{eqnarray}
%\end{widetext}
which can be solved numerically \citep{1989nrpa.book.....P, 2000PhDT.......265K}. Similarly, Eq.~(\ref{ksecondary}) for secondary pairs and Eq.~(\ref{kphoton}) for photons can be solved numerically as well. It should be noted that at the initial moment, i.e., the first time interval, there are no photons in the emitting blob. It means that when solving these three continuity equations numerically, any processes related to photons, including IC scattering, $p\gamma$ process, BH process, synchrotron self-absorption and $\gamma \gamma$ pair annihilation, are not triggered at the initial moment. As time progresses, the number densities of charged particles and photons in the emitting blob increase, enhancing the cooling efficiencies of various processes considered in these three continuity equations. Therefore, these three equations are interdependent and need to be solved simultaneously. After calculating the temporal evolution of $N_{\rm e/p}(\gamma_{\rm e/p}, t)$, $N_{\rm e,sec}(\gamma_{\rm e}, t)$, $n_{\rm soft/\gamma}(E_{\rm soft/\gamma},t)$ numerically, model predicted light curves and the evolution of SEDs can be obtained. In conventional modeling, secondary particle radiation from hadronic interactions typically makes a secondary dominant contribution to the electromagnetic spectrum. Consequently, the IC cooling of primary electrons and secondary pairs is primarily determined by the emission from primary electrons. In this work, we anticipate that some prominent components will be contributed by hadronic processes, which would enhance the cooling of both primary and secondary electrons. Solving the above continuity equations ensures that the obtained results are self-consistent. 

\section{Application to the hard X-ray excess}\label{app}
In this section, we apply the time-dependent leptohadronic model to interpret two observations of hard X-ray excess of Mrk 421. These two hard X-ray excess signatures are found in low-states, so we assume that particles injection is continuous and has a constant rate. In the modeling, we also assume that primary relativistic electrons and protons are injected at the same time. By maintaining the injection of primary particles, the time evolution and steady-state SEDs can be obtained by solving the continuity equation described above.
\subsection{Hard X-ray excess in MJD 56302}
Previous studies have suggested that when hadronic interactions are important, secondary pairs emission may fill the gap between the two bumps of blazars' SED \citep{2015MNRAS.447...36P, 2019MNRAS.483L..12C, 2019NatAs...3...88G, 2022PhRvD.106j3021X}. For the hard X-ray excess observed in Mrk 421, a similar structure has also been detected in PKS 2155-304 \citep{2016ApJ...831..142M, 2017ApJ...850..209G}. Previously, an attempt was made to explain PKS 2155-304's hard X-ray excess using secondary emission from BH process, but this required introducing a highly super-Eddington jet power \citep{2020A&A...639A..42A}. To explore the universality of the leptohadronic model in explaining the hard X-ray excess of Mrk 421 found in 2013 \cite{2016ApJ...827...55K}, we will not introduce the super-Eddington jet power in the modeling below. To simplify the adjustment of free parameters, we boldly assume that the jet power is equal to the Eddington luminosity.

\begin{table*}
\setlength\tabcolsep{1pt}
\caption{Parameters for the SED fitting with the time-dependent leptoharonic model. Columns with serial number: (1) the dominant hadronic processes. (2) the Doppler factor. (3) the magnetic field. (4) the blob radius. (5) the injection luminosity of relativistic electrons in the comoving frame. (6) the electron spectral index before $\gamma_{\rm e,b}$. (7) the electron spectral index after $\gamma_{\rm e,b}$. (8) the break electron Lorentz factor. (9) the proton spectral index. (10) the maximum proton Lorentz factor. (11) the minimum electron Lorentz factor. (12) the maximum electron Lorentz factor. (13) the minimum proton Lorentz factor. (14) the injection luminosity of relativistic protons in the comoving frame. (15) number density of cold protons. (16) the kinetic power in cold protons. (17) the power carried in the magnetic field $P_{\rm B}=\pi R^2c\Gamma^2B^2/8\pi$.}
\centering
\begin{tabular}{cccccccccc}
\hline\hline																			
Free 	&	$\delta_{\rm D}$	&	$B$	&	$R$	&	$L_{\rm e,inj}$	&	$\alpha_{\rm e,1}$	&	$\alpha_{\rm e,2}$	&	$\gamma_{\rm e,b}$	&	$\alpha_{\rm p}$	&	$\gamma_{\rm p,max}$		\\
Parameters	&		&	(G)	&	(cm)	&	$\rm(erg~s^{-1})$	&		&		&		&		&			\\
	(1) & (2) & (3) & (4) & (5) &(6)& (7) & (8) & (9) & (10)  \\
	\hline
$p\gamma$ \& BH	&	13	&	0.18	&	$5\times10^{16}$	&	$5\times10^{41}$	&	1.7	&	4.5	&	$7\times10^{4}$	&	2	&	$2.5\times10^{7}$	\\
$pp$	&	40	&	0.9	&	$1.4\times10^{15}$	&	$1.2\times10^{40}$	&	1.8	&	6.5	&	$3\times10^{4}$	&	1.5	&	$7.1\times10^{4}$	\\
\hline
Fixed/Derived & $\gamma_{\rm e,min}$ & $\gamma_{\rm e,max}$ & $\gamma_{\rm p,min}$ &  $L_{\rm p,inj}$ & $n_{\rm H}$ & $P_{\rm p, cold}$ & $P_{\rm B}$ &  &   \\
Parameters	&		&		&		&	$\rm(erg~s^{-1})$	&	$\rm(cm^{-3})$	&	$\rm(erg~s^{-1})$	&	$\rm(erg~s^{-1})$	&		&			\\
	 & (11) & (12) & (13) & (14) &(15)& (16) & (17)  &   \\
\hline
$p\gamma$ \& BH	&	$3\times10^{2}$	&	$5\times10^6$	&	1	&	$1.1\times10^{45}$	&	0	&0	&	$5.1\times10^{43}$	&		&			\\
$pp$	 &	$3\times10^{2}$	&	$5\times10^6$	&	1	&	$5.1\times10^{43}$	&	$2\times10^{5}$	& $8.2\times10^{46}$ 	&	$9.5\times10^{42}$	&		&	\\
\hline\hline
\label{parameters}
\end{tabular}
\end{table*}

\begin{figure}
\subfloat{
\includegraphics[width=0.55\columnwidth]{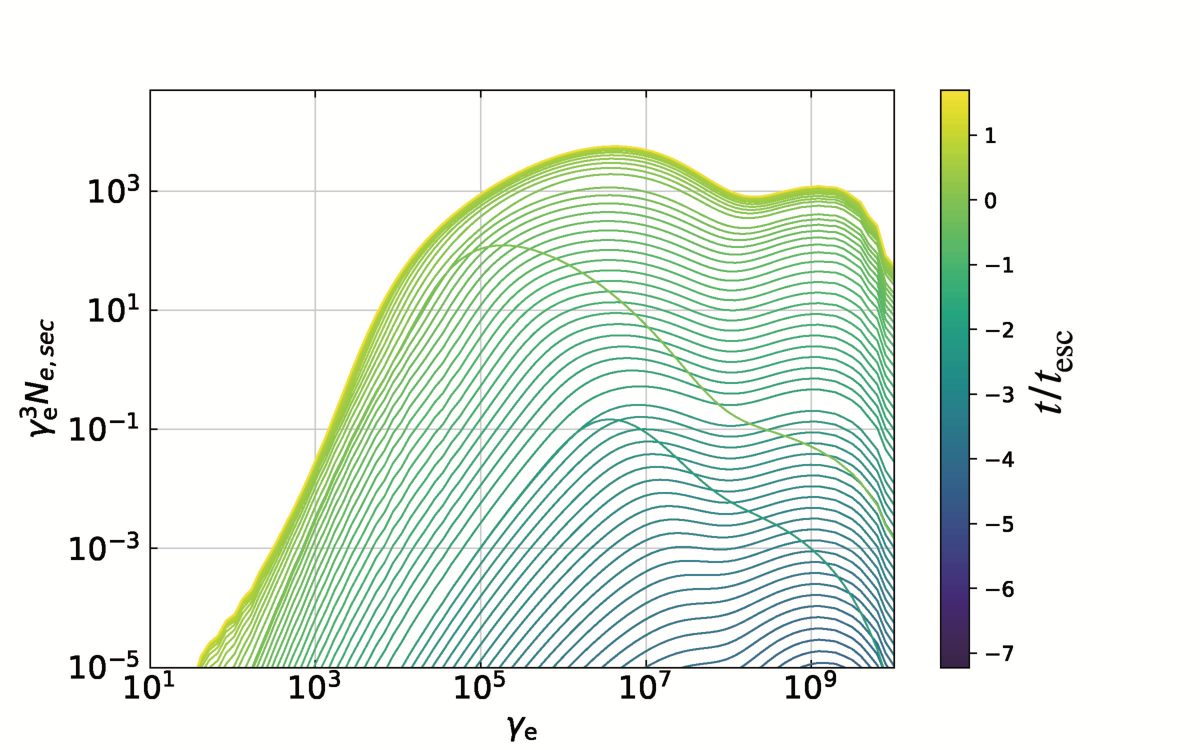}
}
\subfloat{
\includegraphics[width=0.55\columnwidth]{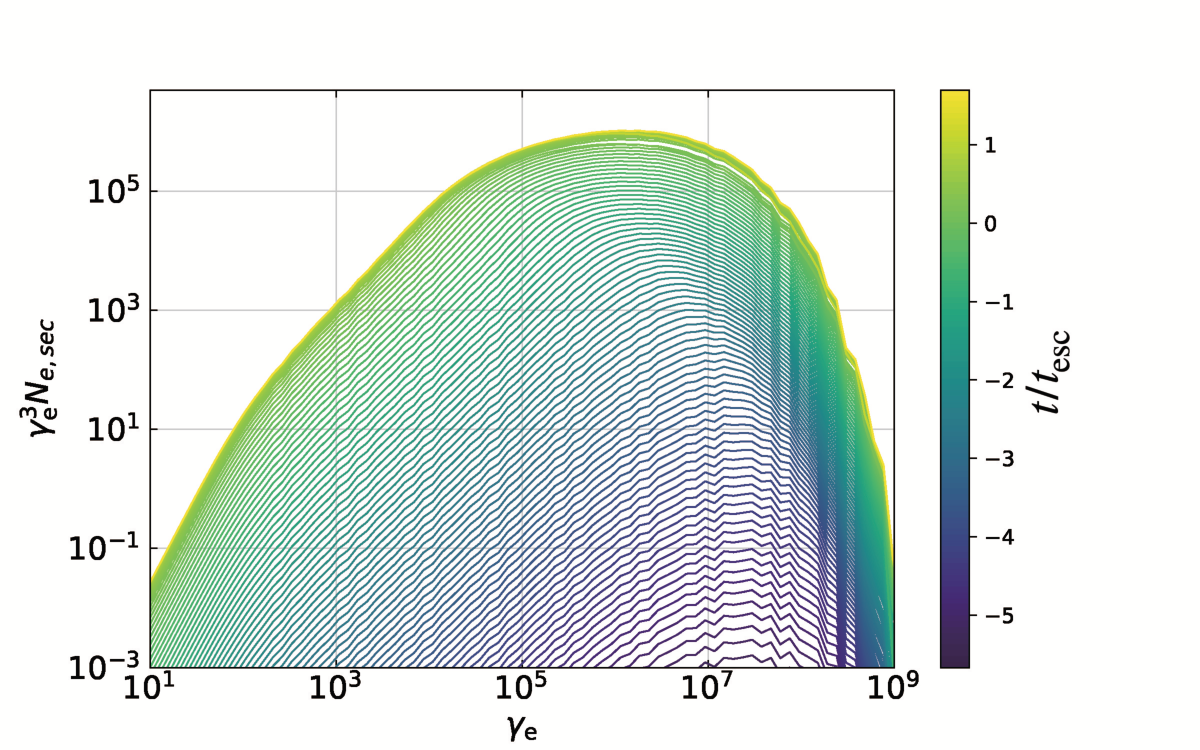}
}
\caption{The temporal evolution of secondary pairs distribution. The left panel show the case of photohadronic interactions, and the right panel show the case of hadronuclear interactions.
\label{EED}}
\end{figure}

\begin{figure*}
\subfloat{
\includegraphics[width=0.55\columnwidth]{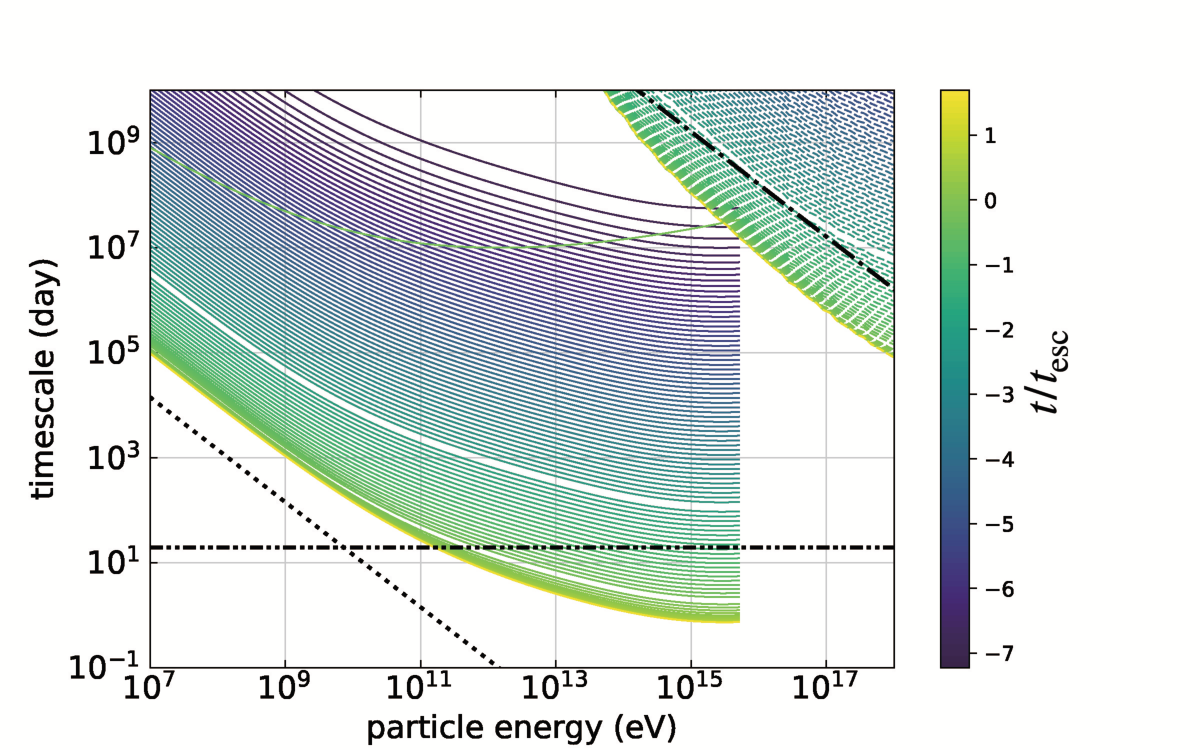}
}
\quad
\subfloat{
\includegraphics[width=0.55\columnwidth]{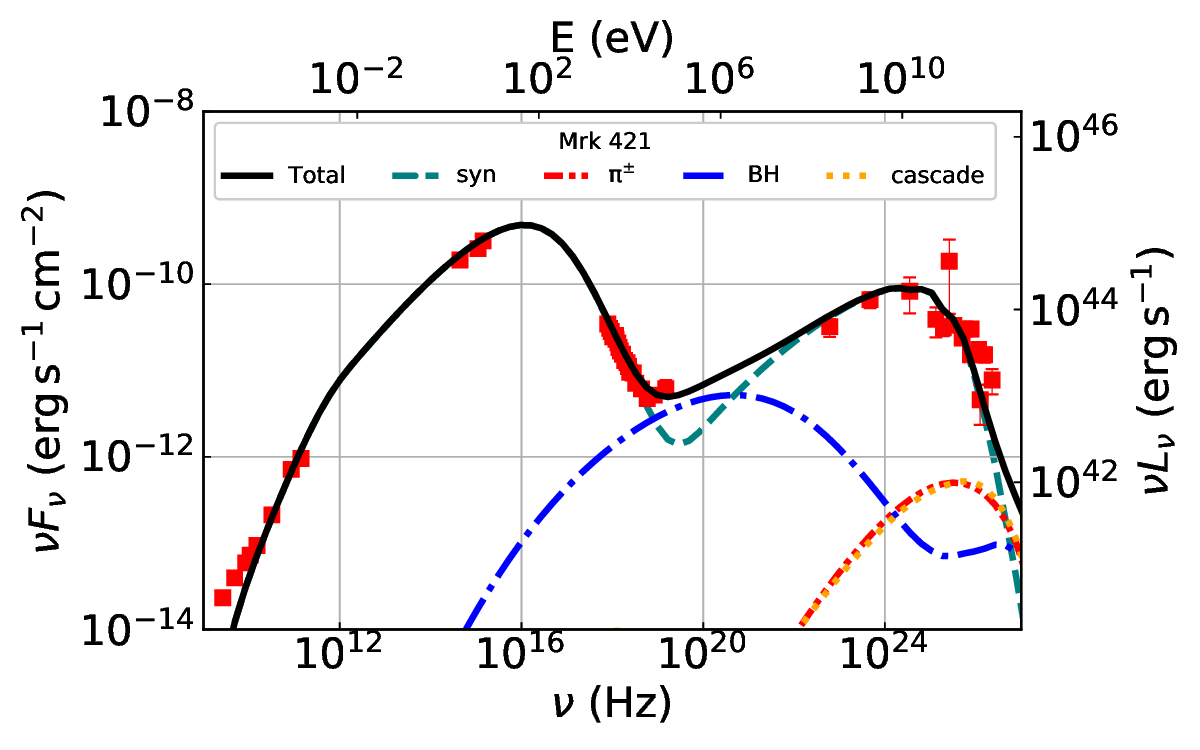}
}
\quad
\subfloat{
\includegraphics[width=0.55\columnwidth]{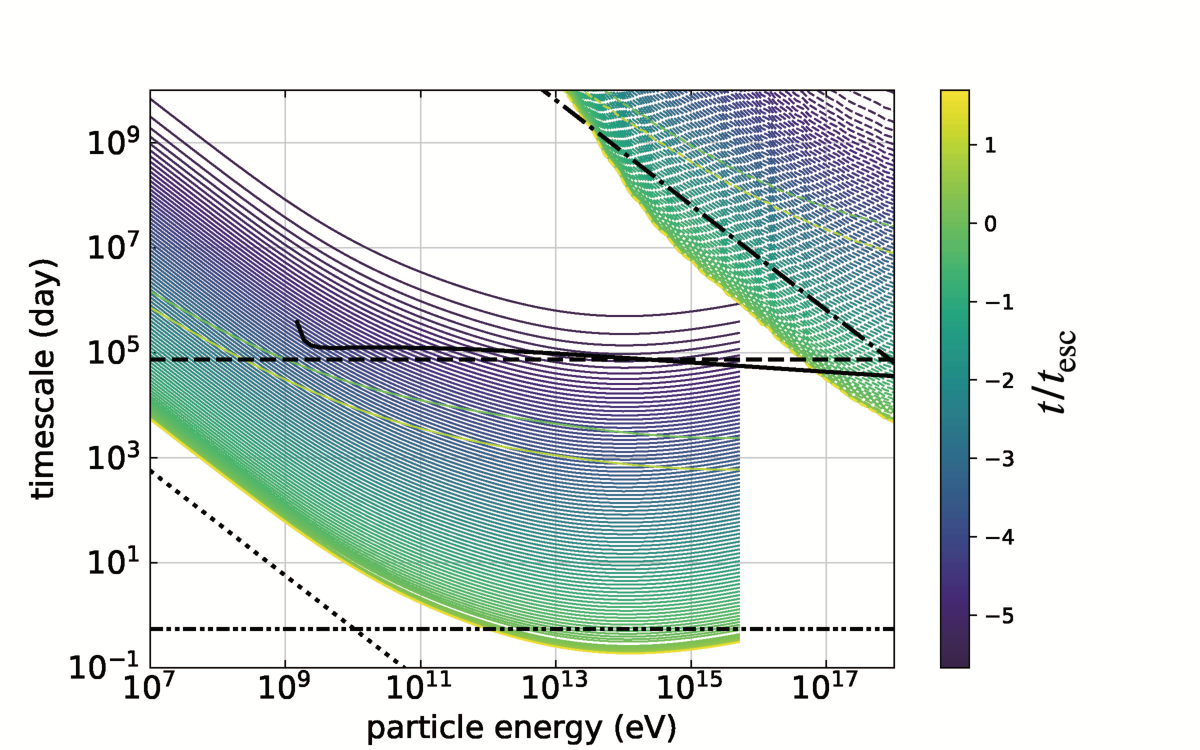}
}
\quad
\subfloat{
\includegraphics[width=0.55\columnwidth]{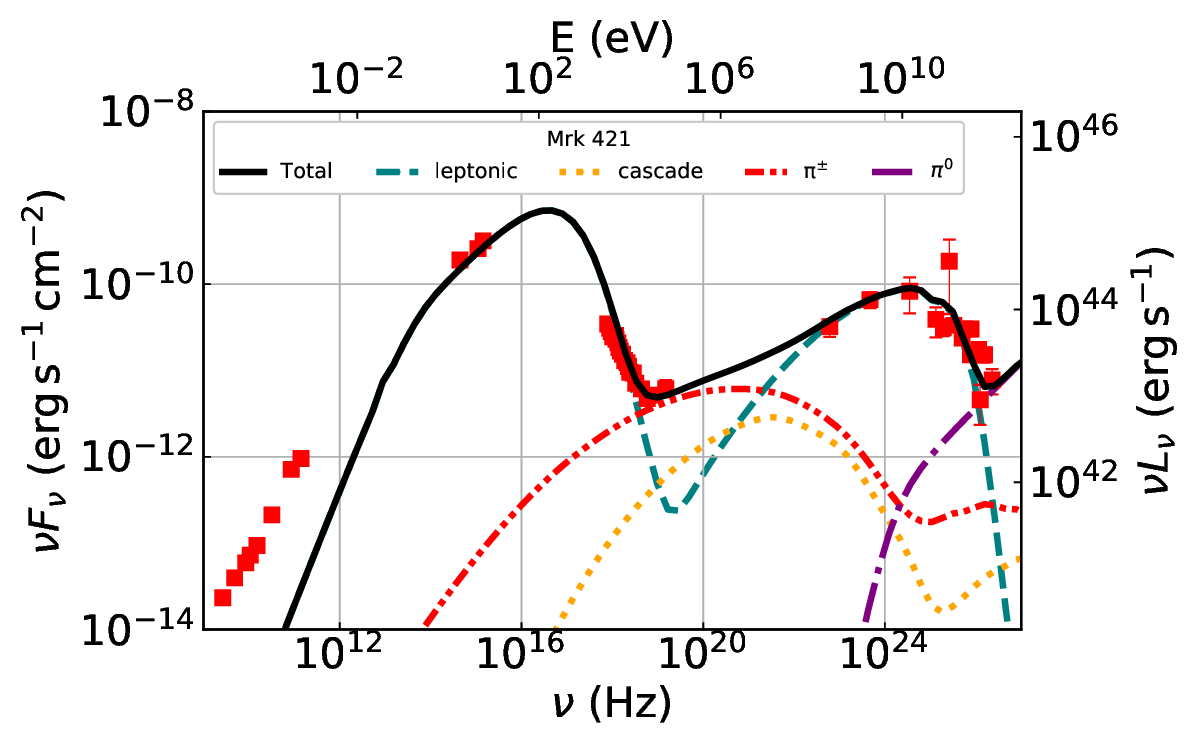}
}
\caption{Leptohadronic SED modelings of Mrk 421. Upper left and lower left panels respectively show the timescales of various processes for relativistic electrons and protons as functions of the particle energy when $p\gamma$ and BH interactions are the dominant hadronic processes and when $pp$ interactions are the dominant hadronic processes. Both the particle energy and timescales are measured in the jet comoving frame. In these two panels, solid and dashed colored curves represent the time evolution of IC cooling and photohadronic (including $p\gamma$ and BH interactions) cooling timescales; black solid, dashed, dotted, dot-dashed, and double dot-dashed curves represent $pp$ cooling, bremsstrahlung cooling, electrons' synchrotron cooling, protons' synchrotron cooling, and escape timescales. Upper right and lower right panels show the fitting results when $p\gamma$ and BH interactions are the dominant hadronic processes, and when $pp$ interactions are the dominant hadronic processes, respectively. Red data points are simultaneous data taken from \citep{2016ApJ...827...55K, 2016ApJ...819..156B}. Teal dashed curve shows the synchrotron and SSC emission from primary electrons. Orange dotted and double dot-dashed curves represent synchrotron emission from secondary electrons from $\pi^\pm$ decay and internal $\gamma \gamma$ annihilation. Blue dot-dashed curve in the upper panel and purple dot-dashed curve in the lower panel are synchrotron emission from secondary electrons from BH process and $\gamma$-ray emission from $\pi^0$ decay.
\label{excess}}
\end{figure*}

In calculations, we adjust the number density of cold protons $n_{\rm H}$ to alter the dominance level of photohadronic interactions (including $p\gamma$ and BH interactions) and $pp$ interactions. When considering photohadronic interactions are dominant, we set $n_{\rm H}=0~\rm cm^{-3}$ and $\Gamma^2L_{\rm p, inj}\simeq L_{\rm Edd}$, where $L_{\rm Edd}=1.26\times10^{38}\rm~erg~s^{-1}\it{M}_{\rm bh}/M_{\odot}$ is the Eddington luminosity, and we take $M_{\rm bh}\simeq1.3\times10^9~M_{\odot}$ for the black hole mass of Mrk 421 in units of the solar mass $M_{\odot}$ \citep{2002A&A...389..742W}. In this scenario, emissions of secondary pairs from both of $p\gamma$ and BH interactions may contribute to the hard X-ray excess. To produce the hard X-ray excess around $20~\rm keV$ through synchrotron emission, the energy of corresponding pairs $E_{\rm e^{\pm}}$ in the comoving frame is $\sim 185\rm~GeV(\frac{1~\rm G}{\it{B}})^{1/2}(\frac{10}{\delta_{\rm D}})^{1/2}$. In BH process, pairs with this energy are generated by protons with energy $E_{\rm p}\approx E_{\rm e^{\pm}}\frac{m_{\rm p}c^2}{m_{\rm e}c^2}\approx 340\rm~TeV$ \citep{2008PhRvD..78c4013K}, and the energy of corresponding target photons is $E_{\rm t}^{\rm BH}\approx 0.01~\rm GeV^2/E_{\rm p}\approx 28~\rm eV$; while in $p\gamma$ interactions, the parent proton energy is $E_{\rm p}\approx 20E_{\rm e^{\pm}}\approx 3.7\rm~TeV$, and the energy of corresponding target photons is $E_{\rm t}^{p\gamma}\approx 0.3\rm~GeV^2/E_{\rm p}\approx81~\rm keV$.
As their energy is extracted from the injected relativistic protons, the fluxes of secondary emissions from $p\gamma$ and BH interactions are determined by their interaction efficiencies. The relation of the BH interaction efficiency $f_{\rm BH}\approx 7\times10^{-31}~\rm cm^2\it{n}_{\rm t}^{\rm BH}\it{R}$, where $7\times10^{-31}~\rm cm^2$ represents the BH cross section weighted by inelasticity, and the $p\gamma$ interaction efficiency $f_{p\gamma}\approx 10^{-28}~\rm cm^2\it{n}_{\rm t}^{p\gamma}\it{R}$, where $10^{-28}~\rm cm^2$ represents the $p\gamma$ cross section weighted by inelasticity, is $f_{\rm BH}/f_{p\gamma}\approx n_{\rm t}^{\rm BH}(E_{\rm t}^{\rm BH})/(143~n_{\rm t}^{p\gamma}(E_{\rm t}^{p\gamma}))$. From the low-energy bump of Mrk 421, it is obvious that the ratio of $n_{\rm t}^{\rm BH}(28~\rm eV)$ to $n_{\rm t}^{p\gamma}(81~\rm keV)$ is much larger than 143. Therefore, it is expected that synchrotron emission of secondary pairs from BH interactions interpret the hard X-ray excess.
%In this scenario, emissions of secondary pairs from these two interactions contribute significantly to the SED in different energy ranges, because the energy of secondary pairs from BH interactions is $E_{\rm e^{\pm}}^{\rm BH}\approx E_{\rm p}\frac{m_{\rm e}c^2}{m_{\rm p}c^2}$ \citep{2008PhRvD..78c4013K}, where $E_{\rm p}$ is the proton energy in the comoving frame, and the energy of secondary pairs from $\pi^\pm$ decay in $p\gamma$ interactions is $E_{\rm e^{\pm}}^{p\gamma}\approx 0.05E_{\rm p}$. In addition, these interactions have the same target photon field, i.e., low-energy bump of Mrk 421. From the perspective of $\delta$-approximation for their cross sections, the energy of target photons $E_{\rm t}^{p\gamma}$ and the energy of protons $E_{\rm p}$ satisfy the relation $E_{\rm t}^{p\gamma}E_{\rm p}\approx0.3~\rm GeV^2$ for the $p\gamma$ interactions, while the same protons interact with photons of energy $E_{\rm t}^{\rm BH}\approx 10~\rm MeV/\gamma_{\rm p}$ in BH interactions. Energies of these two target photons have the relation $E_{\rm t}^{\rm BH}/E_{\rm t}^{p\gamma}\approx1/30$. 
When considering $pp$ interactions are non-negligible, we set $\Gamma^2L_{\rm p, inj}\simeq P_{\rm p, cold} \simeq 0.5L_{\rm Edd}$ as suggested in our previous work \citep{2022A&A...659A.184L, 2022PhRvD.106j3021X}, where $P_{\rm p, cold}=\pi R^2c\Gamma^2m_{\rm p}c^2n_{\rm H}$ represents the kinetic power in cold protons. It is expected that secondary pairs from $\pi^\pm$ decay mainly contribute to the hard X-ray excess. With parameters listed in TABLE~\ref{parameters}, temporal evolutions of secondary pairs from photohadronic and $pp$ interactions are given in FIG.~\ref{EED}. From TABLE~\ref{parameters}, it can be found that the total jet power is approximately equal to the Eddington luminosity as the parameter settings we established before. Power carried in magnetic field and relativistic electrons is insignificant. The ratio of magnetic power to proton power and the ratio of electron power to proton power are $\sim10^{-5}$--$10^{-4}$ and (2--5)$\times10^{-3}$, respectively, which are similar to some previous leptohadronic modeling studies \citep{2013ApJ...768...54B, 2015MNRAS.448..910C}. In the left panel of FIG.~\ref{EED}, it can be seen that the major bump of pairs distribution originated from BH interactions peaks around $\gamma_{\rm e, peak}\sim3\times10^6$, leading to an expected synchrotron peak frequency $\nu_{\rm syn}\simeq 3.3\times10^{19}~\rm Hz(\frac{\gamma_{e, peak}}{3\times10^6})^2(\frac{\it{B}}{0.1~\rm G})(\frac{\delta_{\rm D}}{10})$, and a second lower bump originated from $p\gamma$ interactions peaks around $\sim3\times10^9$. In the right panel of FIG.~\ref{EED}, it can be seen that pairs distribution originated from $\pi^\pm$ decay peaks around $\sim10^6$. Therefore, with pairs' having the Lorentz factor of about $10^6$, their synchrotron emission peaks in the hard X-ray band. The corresponding various timescales and modeling results are shown in FIG.~\ref{excess}. In the left two panels (timescales), it can be seen that relativistic electrons primarily cool through synchrotron radiation, especially for high-energy secondary pairs, where the severe KN effect weaken the IC cooling. As for the cooling of relativistic protons,  the number density of soft photons increases over time, and the efficiency of the $p\gamma$ and BH processes surpasses that of the synchrotron cooling gradually. When the number density of cold protons is large, $pp$ interactions become the most important cooling process for relativistic protons. Additionally, one may be concerned about if bremsstrahlung cooling of electrons becomes important. With the current parameters, we neglect bremsstrahlung cooling because its cooling timescale \cite{1964ocr..book.....G} $\sim7.3\times10^4(n_{\rm H}/2\times10^5~\rm cm^{-3})~day$ is much longer than those of synchrotron and IC cooling. In the right two panels (modeling results), it can be seen from the upper panel that, the hard X-ray excess is interpreted by the emission from BH pairs as expected. Furthermore, the hard X-ray band is the sole contribution of hadronic processes. Currently, observations of the hard X-ray excess in Mrk 421 have been made during low states \citep{2016ApJ...827...55K,2021MNRAS.504.1427A}. If future observations during flare states are possible, if there is a correlation between the hard X-ray band and other band variabilities could be used to test this model result. In the lower right panel, the hard X-ray excess is interpreted by the emission of secondary pairs from $\pi^\pm$ decay in $pp$ interactions. In addition, $\pi^0$ decay $\gamma$-ray also has contribution at VHE band, predicting a hard TeV spectrum at higher energies, even though this hard spectrum has not yet been observed. For $pp$ interactions, the contribution of $\pi^0$ decay in the VHE band is unavoidable, as the flux in the VHE band is comparable to that in the hard X-ray excess. In our modeling, a hard proton injection spectrum ($\alpha_{\rm p}=1.5$) is introduced because the secondary particle distribution from the $pp$ process essentially follows the distribution of primary protons. If the proton spectrum is relatively soft ($\alpha_{\rm p}\gtrsim 2$), the secondary radiation is also expected to be a flat or soft spectrum, which is not conducive to explaining the hard X-ray excess. Note that in obtaining the fitting results for the lower right panel, the $p\gamma$ and BH interactions are also calculated. However, their contributions to the SED are almost negligible because protons' maximum energy is set relatively low ($6.7\times10^{13}\rm~eV$), resulting in much lower interaction efficiencies compared to the $pp$ interactions. From the above results, we show that without introducing extreme physical parameters, such as the super-Eddington jet power, both photohadronic and $pp$ processes can explain the hard X-ray excess of Mrk 421 found in 2013. However, their variability signatures differ. For the photohadronic processes, the contribution of the BH process to hard X-ray band strongly depends on the evolution of target photons with energy $\sim28\rm~eV$ produced by primary electrons' synchrotron emission. It means that the hard X-ray radiation originating from the BH process will be delayed compared to the variability of $\sim28\rm~eV$ emission. The delay timescale approximately corresponds to the cooling timescale $T_{\rm cool}^{\rm obs}$ of the electrons in the observers' frame, i.e., $T_{\rm cool}^{\rm obs}\approx 1.6\rm~h(\frac{1\rm~G}{\it{B}})^{1.5}(\frac{10}{\delta_{\rm D}})^{0.5}$. In contrast, for the $pp$ process, the secondary pairs from $\pi^\pm$ decay have no connection with soft photons, and thus there is no such delay. In future variability observations, the presence of delayed variability in the hard X-ray band could serve as a test to distinguish between photohadronic and $pp$ origins.

\subsection{Hard X-ray excess in MJD 57422-57429}
Recently, an additional spectral component beyond the low-energy bump of Mrk 421 is found at low activity state in 2016 \citep{2021MNRAS.504.1427A}. This structure is similar to the hard X-ray excess in 2013, but its relative flux compared to the low-energy bump is higher. This implies that this structure is more pronounced and more difficult to explain compared to the hard X-ray excess in 2013. 

\begin{figure}
\subfloat{
\includegraphics[width=0.5\columnwidth]{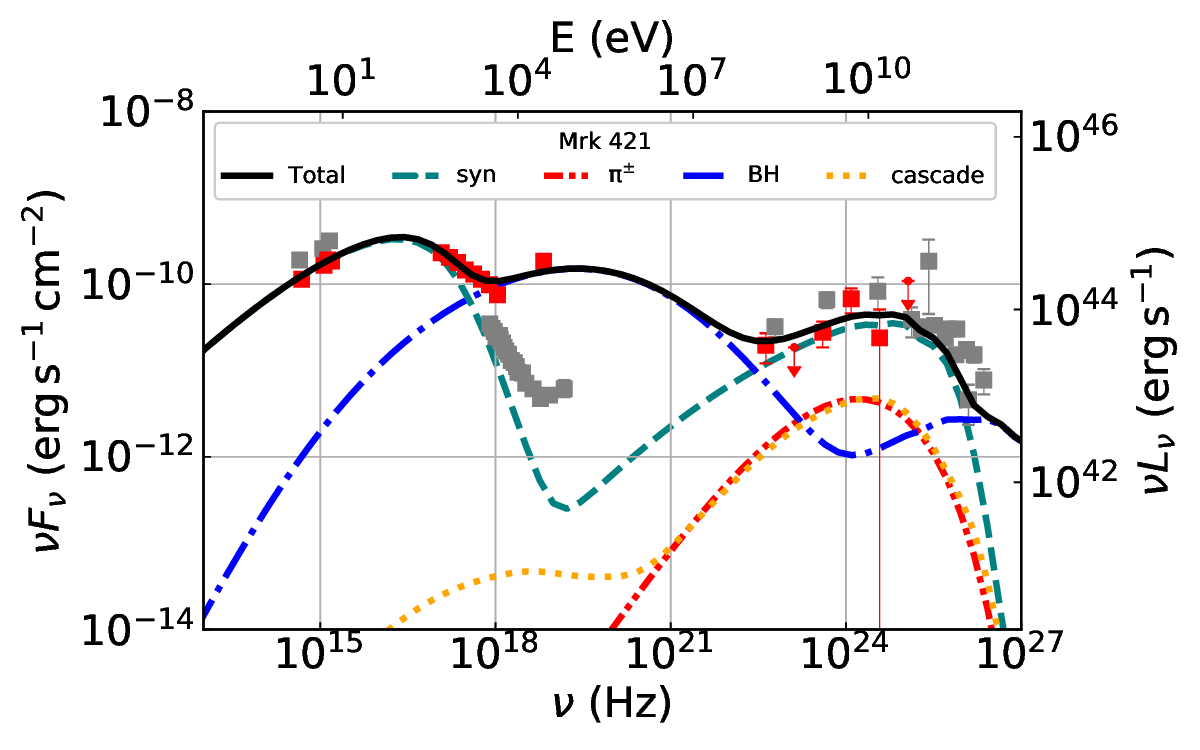}
}
\quad
\subfloat{
\includegraphics[width=0.5\columnwidth]{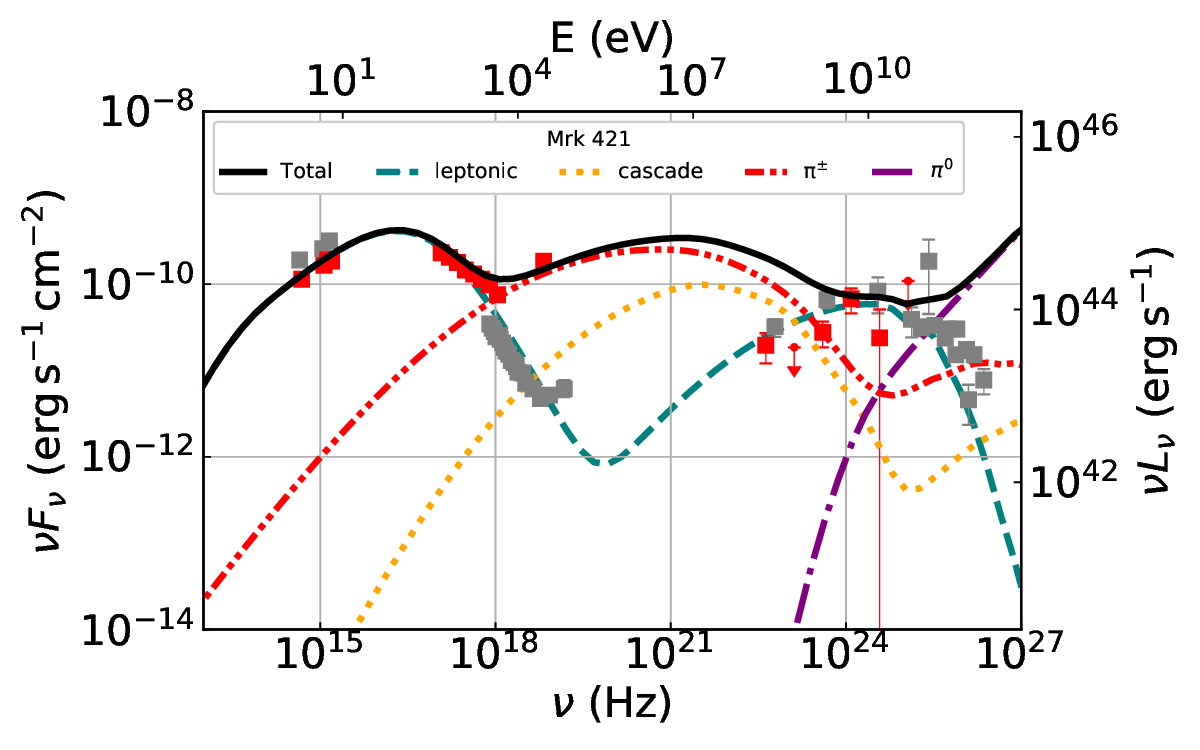}
}
\quad
\subfloat{
\includegraphics[width=0.5\columnwidth]{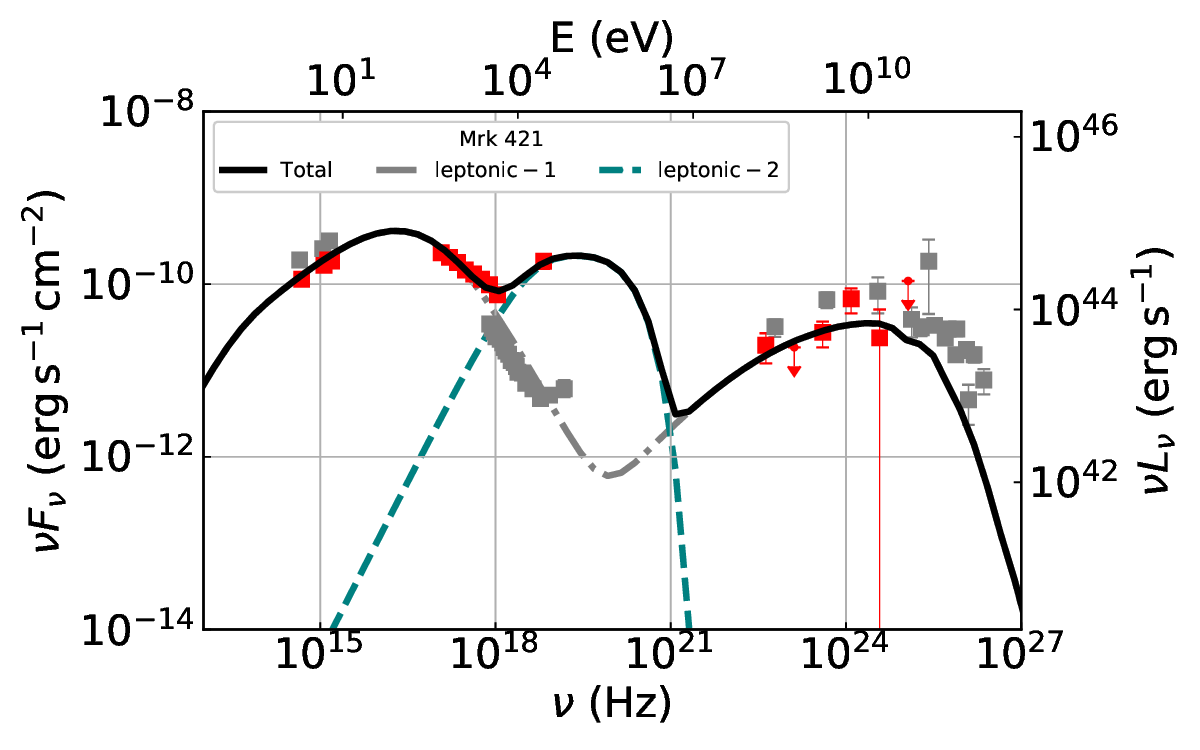}
}
\caption{SED modeling of Mrk 421. Red data points from the optical to hard X-ray range are taken from \cite{2021MNRAS.504.1427A} during the time interval MJD 57422--57429, while the red data points in the $\gamma$-ray band are obtained over the same time interval. Gray data points correspond to the red data points shown in FIG.~\ref{excess}. Upper left panel shows the fitting result that $p\gamma$ and BH interactions are the dominant hadronic processes, upper right panel shows the fitting result that $pp$ interactions are the dominant hadronic processes, and lower panel shows the fitting result of the two-zone leptonic model, respectively. The meaning of all curves are explained in the inset legend.
\label{two}}
\end{figure}

In \citep{2021MNRAS.504.1427A}, only the simultaneous data from optical to hard X-ray bands is provided during the time interval MJD 57422--57429, lacking data for the high-energy bump. To better constrain the origin of the newly discovered hard X-ray component, we obtain $Fermi$ GeV data within the same time interval. The region of interest (ROI) centred on the source's `RA' and `Dec.' is set to $12^\circ$. Below 100 MeV, the effective area of the LAT rapidly decreases, and the point-spread function increases to above $\sim6^\circ$. To prevent the point source analysis from being challenging, we follow the standard data selection recommendations and restrict the photon energy range to 100 MeV--300 GeV. The tool used for data analysis is easyFermi \citep{2022A&C....4000609D}, which provides an open-source, user-friendly graphical interface for intermediate analysis of $Fermi$-LAT data within the Fermipy framework \cite{2017ICRC...35..824W}. The model file was generated based on Fermi-LAT fourth source catalogue Data Release 3 (4FGL-DR3; \citep{2022ApJS..260...53A}), which includes point sources within the ROI and standard Galactic (gll\_iem\_v07) and the isotropic (iso\_P8R3\_SOURCE\_V3\_v1) diffuse emission components. To reduce the contamination from Earth limb $\gamma$-ray, we apply a maximum zenith angle cut of greater than $90^{\circ}$. For the energy spectrum, we further set the upper energy limit to 84.1 GeV (corresponding to the maximum photon energy within the ROI) and divide the energy range of 0.1--84.1 GeV into six equal logarithmic energy bins. For data points with TS less than 9, we treat them as 95 per cent confidence level upper limits. The obtained $Fermi$ data is shown in FIG.~\ref{two}. It is obvious that the hard X-ray component cannot be simply extended to connect with the $Fermi$ data. This suggests that the hard X-ray component is a new radiation component. 

Following the main purpose of this work, we first check whether emissions from photohadronic and hadronuclear interactions can explain the new hard X-ray component. Based on the parameters in Table~\ref{parameters}, by adjusting individual parameters, we obtained the fitting results dominated by photohadronic ($\alpha_{\rm e,2}=6$, $\gamma_{\rm e,b}=10^5$, $L_{\rm e,inj}=3\times10^{41}\rm~erg/s$, $\gamma_{\rm p,max}=5\times10^6$, $L_{\rm p,inj}=150\times L_{\rm Edd}/\delta_{\rm D}^2\rm~erg/s$) and hadronuclear interactions ($B=0.7~\rm G$, $\alpha_{\rm e,2}=4.5$, $\gamma_{\rm e,b}=2\times10^4$, $\alpha_{\rm p}=1.3$, $L_{\rm p,inj}=17.5\times L_{\rm Edd}/\delta_{\rm D}^2\rm~erg/s$), respectively, as shown in the upper left and upper right panels of FIG.~\ref{two}. It can be seen that secondary emission from BH process provides an acceptable fitting to the hard X-ray excess. Although a super-Eddington jet power is required due to the relative high flux of the hard X-ray component. For $pp$ interactions, it failed to interpret the SED, since the broad distribution of secondary electrons leads to an overshoot of the $Fermi$ data. In the lower panel of FIG.~\ref{two}, we also propose a two-zone leptonic model. It can be seen that the synchrotron radiation from electrons with a narrow distribution ($5\times10^5\leqslant \gamma_{\rm e}\leqslant 5\times10^6$) in the second region is responsible for the hard X-ray component. This solution is generally consistent with previous explanations for the associated flares of X-ray and VHE. The difference is that this hard X-ray component was discovered during a low state, indicating that the energetic narrow distribution of electrons is accelerated in a larger-scale radiation region. Shear acceleration might provide a possible explanation \citep{2017ApJ...842...39L, 2020Natur.582..356H, 2021MNRAS.505.1334W}.

\section{Application to correlation between X-ray and VHE flares}\label{correlation}
An intriguing VHE flare and a possible corresponding X-ray flare with spectral hardening of Mrk 421 were detected on MJD 57788 \citep{2021A&A...655A..89M}. The observed VHE variability timescale $t_{\rm var}$ is $\sim 2\rm~hours$, suggesting a compact emitting zone, i.e., $R\approx 2.2\times10^{15}\rm~cm (\frac{\it{t}_{\rm var}}{2~\rm h})(\frac{\delta_{\rm D}}{10}$). In the optical--UV and GeV bands, no significant variability is observed, indicating that the detected VHE and X-ray flares may be attributed to a more energetic and narrower particle energy distribution. In the previous section, we have shown in FIG.~\ref{EED} that hadronic interactions also can naturally generate energetic pairs. In the following, our purpose is to test if hadronic interactions can be used to interpret the correlated X-ray and VHE flare in MJD 57788.

\begin{figure}
\subfloat{
\includegraphics[width=0.5\columnwidth]{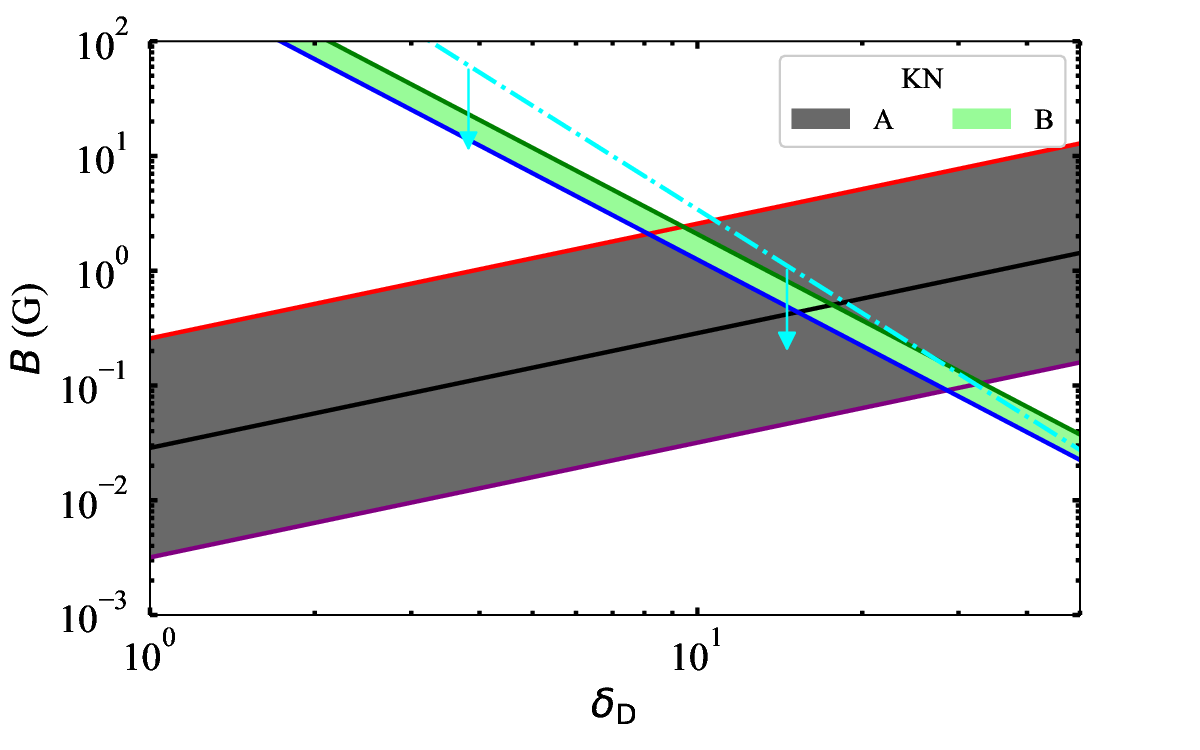}
}
\subfloat{
\includegraphics[width=0.55\columnwidth]{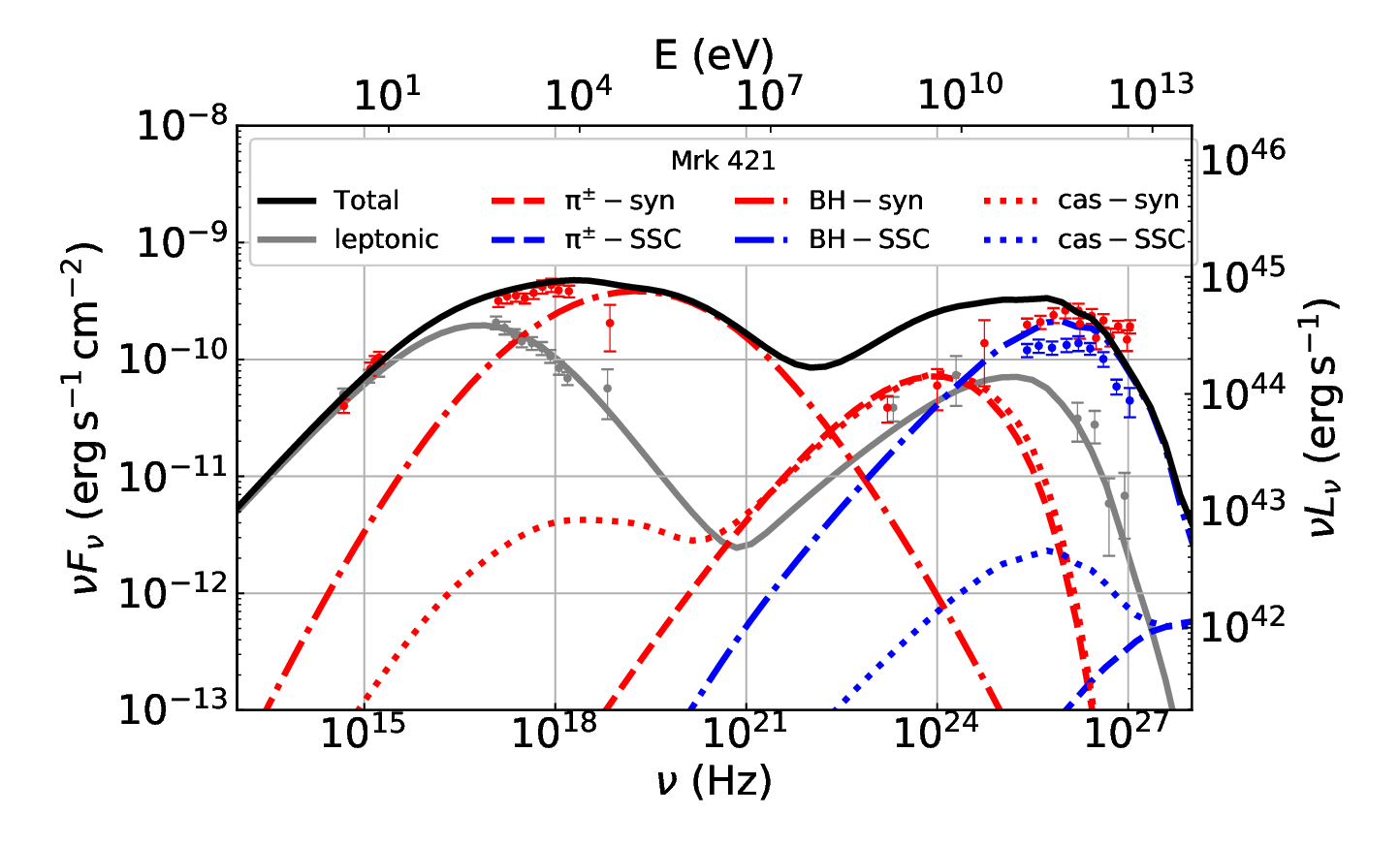}
}
\caption{Left panel: The parameter space of Mrk 421 in the framework of one-zone SSC model under the KN regime. The grey (A) and palegreen (B) regions represent the constraints of peak frequencies and peak luminosities of the double bumps, respectively. In region A, red, black, and purple lines are obtained at the high-energy peak frequencies at $6\times 10^{24}~\rm Hz$, $1.8\times 10^{25}~\rm Hz$, and $5.4\times 10^{25}~\rm Hz$, respectively. In region B, blue and green lines are calculated based on the observed variabilities of 2.5 hour and 1.5 hour, respectively. The cyan dash-dotted line with arrows represents the parameter space that the severe KN effect is trigged. It can be seen that the high-energy peak from SSC emission is significantly affected by the KN effect. Right panel: The one-zone leptohadronic modeling of the simultaneous SED of Mrk 421. Red and gray data points represent the simultaneous SED of flare and pre-flare states in MJD 57788 and 57786 \citep{2021A&A...655A..89M}. The meaning of all curves is explained in the inset legend. The parameters adopted for the emitting region and protons are $B=0.4~\rm G$, $R=2.3\times10^{15}~\rm cm$, $\delta_{\rm D}=14$, $\alpha_{\rm p}=2$, $\gamma_{\rm p, max}=2.5\times10^6$, $L_{\rm p,inj}=4.5\times10^{47}\rm~erg/s$.
\label{space}}
\end{figure}

In MAGIC Collaboration et al. \cite{2021A&A...655A..89M}, simultaneous multiwavelength SEDs of pre-flare (MJD 57786) and flare states (MJD 57788), and multiwavelength light curves around the flare state are provided. 
%It is necessary to figure out if SEDs of these two states could originate from the same emitting region. 
For the modeling of pre-flare state, the classic one-zone SSC model is defaulted as usual. Here we apply the analytical method proposed in our recent work to obtain the space of physical parameters, including $B$, $R$, and $\delta_{\rm D}$ \citep{2024MNRAS.528.7587H}. With the constraints of peak frequencies and peak luminosities of two bumps, the obtained parameter space in the KN regime is shown in the left panel of FIG.~\ref{space}. Within the obtained parameter space, we utilize the iminuit\footnote{https://ascl.net/2108.024} \citep{2021ascl.soft08024D} to search for permissible physical parameter combinations that allow the secondary emission from $pp$ or photohadronic interactions, combined with pre-flare leptonic emission, to fit the flare state SED. For the secondary emission from $pp$ interactions, it is impossible to get a solution. As expected, the synchrotron emission of secondary pairs from $\pi^\pm$ decay and the $\gamma$-ray emission from $\pi^0$ decay respectively explain the hard X-ray and VHE flares. The branching ratio into $\pi^\pm$ and $\pi^0$ is 1/6 and 1/3, respectively. Even assuming secondary pairs are in fast cooling regime, it is impossible for the emission from secondary pairs to exceed the flux of $\pi^0$ decay $\gamma$-ray. This contradicts the fact that the observed flux ratio of X-ray and VHE spectra in the flare state exceeds unity. Next, we check if secondary emission from $p\gamma$ and BH interactions could provide a possible explanation. As shown in the upper panel of FIG.~\ref{excess}, by decreasing the maximum proton Lorentz factor, it is expected that synchrotron and IC emission of secondary pairs from BH interactions would trigger X-ray and VHE flares, while the secondary pairs from $p\gamma$ interactions mainly contribute at $\sim$GeV band with a lower flux. Unfortunately, within the parameter space indicated by the pre-flare SED fitting with the one-zone SSC model, iminuit was unable to find a valid fitting parameter combination. As shown in the right panel of FIG.~\ref{space}, the primary reason is that when the secondary pair synchrotron emission from BH interactions contributes to hard X-rays, secondary pair synchrotron emission from $\pi^\pm$ decay overshoots the GeV data, which contradicts the non-detection of significant GeV variability. Note that the dominant peak of cascade synchrotron emission is quite similar to that of secondary pair synchrotron emission from $\pi^\pm$ decay. It is because that $\pi^0$ decay $\gamma$-ray is dominant in the $\gamma \gamma$ annihilation, and luminosity branching ratios into $\pi^0$, $\pi^\pm$ and energies of secondary pairs generated in $\gamma \gamma$ annihilation and $\pi^\pm$ decay are similar to each other. In the one-zone scenario, reconciling this issue is challenging because the low-energy bump of Mrk 421 pre-flare state has already been fixed. Even considering that the hadronic processes and the pre-flare emission are not in the same emission region, this issue seems unresolved. Since the target photon energy is $\sim~\rm1~keV$ and the flux in the flare state is similar to that in the pre-flare state, the emission from BH process would still enhance the efficiency of the $p\gamma$ interactions, causing it to continue to overshoot the GeV data.

\section{Discussion and conclusion}\label{DC}
In this work, we investigate if emissions from hadronic interactions could provide a possible interpretation of the hard X-ray excess and correlated flares between X-ray and VHE bands. In this work, we employ the time-dependent model to ensure the accuracy of our calculations, especially in the case of strong cascade radiation. By considering a strategy that utilizes smaller time bins across multiple iterations, we can more comprehensively account for the contribution of cascade processes in our fitting. This approach allows the cascade radiation to continuously engage in photohadronic interactions, IC processes and subsequent cascade processes. For the hard X-ray excess found in 2013 \cite{2016ApJ...827...55K}, our results suggest that the emission of secondary pairs from BH interactions and the emission of secondary pairs from $\pi^\pm$ in $pp$ interactions can both interpret the hard X-ray excess without introducing a super-Eddington jet power. For the hard X-ray excess found in 2016 \cite{2021MNRAS.504.1427A}, we suggest that the emission of secondary pairs from BH interactions could provide a possible explanation, but a super-Eddington jet power is inevitably needed. Therefore, a two-zone leptonic model is applied as well. For correlated flares between X-ray and VHE bands found in MAGIC Collaboration et al. \cite{2021A&A...655A..89M}, due to the constraint of low-energy GeV data, hadronic emissions fail to provide a possible interpretation. Therefore, another electron population with a narrow energetic distribution may still be necessary \citep{2015A&A...578A..22A, 2021A&A...655A..89M}, implying that the SED of 421 is contributed by multiple emitting zones.

In our time-dependent modeling, steady-state particle spectra are derived to fit the average SEDs. The acceleration process is not introduced in our current model, only an injection term is introduced as an alternative to energy injection. Nonetheless, based on the results of the fitting, we can still provide constraints on the underlying acceleration mechanism. For example, if the lepton model is used to explain the SEDs, a second blob of electrons with a narrow energy range is required for both MJD 57422-57429 and MJD 57788 \citep{2021A&A...655A..89M}. However, the first case corresponds to a condition where all bands remain in a steady state over an extended period \citep{2021MNRAS.504.1427A}, unlike the second case,  which corresponds to a condition where the VHE bands show variability on multi-hour timescales \citep{2021A&A...655A..89M}. Consequently, the dissipation region associated with these two cases differ significantly in size. In the first case, electrons potentially gain energy through shear acceleration within large-scale jets \citep{2017ApJ...842...39L}, a process that has been extensively studied \citep{2021MNRAS.505.1334W,2022ApJ...933..149R,2023ApJ...958..169W} and used to explain the extended X-ray and VHE emission on kpc scales \citep{2020Natur.582..356H,2023MNRAS.524...76Z,2023MNRAS.525.5298H}. In the second case, the electrons are accelerated to high energies within a compact region where shear acceleration is inefficient. Hence, alternative mechanisms, such as a pileup in the electron energy distribution due to the stochastic acceleration \citep{2020A&A...637A..86M}, are likely to be involved. Introducing the acceleration process into the time-dependent model and fitting the multi-wavelength light curve has the potential to improve our understanding of the acceleration and radiation mechanisms in the jet. Therefore, in a forthcoming paper, we will carry out more comprehensive studies with the multi-wavelength light curve using our time-dependent models.

\backmatter

\bmhead{Acknowledgements}
We thank the anonymous referee for constructive comments and suggestions. This work is supported by the National Key R\&D Program of China (2023YFB4503300), and the National Natural Science Foundation of China (NSFC) under the grants No. 12203043 and No. 12203024.

\bibliography{sn-bibliography}% common bib file

\end{document}